\documentclass{aastex}

\usepackage{graphicx}	
\usepackage{amsmath}	
\usepackage{amssymb}	
\usepackage{color}
\usepackage{ulem}

\title{An Adaptive Homomorphic Aperture Photometry Algorithm for Merging Galaxies}
\author{J. C. Huang and C. Y. Hwang}
\affil{Graduate Institute of Astronomy, National Central University, Jungli, Taoyuan 32001, Taiwan}
\email{Huang: m969003@astro.ncu.edu.tw, Hwang: hwangcy@astro.ncu.edu.tw}

\begin{document}

\begin{abstract}
We present a novel automatic adaptive aperture photometry algorithm for measuring the total magnitudes of merging galaxies with irregular shapes.  First, we use a morphological pattern recognition routine for identifying the shape of an irregular source in a background-subtracted image.  Then, we extend the shape of the source by using the {\it Dilation} image operation to obtain an aperture that is quasi-homomorphic to the shape of the irregular source.  The magnitude measured from the homomorphic aperture would thus have minimal contamination from the nearby background.  As a test of our algorithm, we applied our technique to the merging galaxies observed by the Sloan Digital Sky Survey (SDSS) and the Canada-France-Hawaii Telescope (CFHT).  Our results suggest that the adaptive homomorphic aperture algorithm can be very useful for investigating extended sources with irregular shapes and sources in crowded regions.
\end{abstract}

\keywords{techniques: image processing --- galaxies: photometry --- galaxies: interactions}

\section{INTRODUCTION}\label{secint}

An important issue in astronomical CCD photometry is obtaining an unbiased flux estimate for an extended source.  Usually, a circular or elliptic aperture is used for estimating the total flux of a source. While the aperture must be sufficiently wide for including most of the fluxes from the source, its size should be limited in order to minimize the sky noise and avoid contamination from the nearby sources.  Some adaptive schemes, such as Kron aperture \citep[]{1980ApJS...43..305K} and Petrosian aperture \citep[]{1976ApJ...209L...1P}, have been proposed for optimization of results.

However, galaxy photometry is not a straightforward process.  For galaxies in a crowded region or for irregular galaxies with odd shapes, it is difficult to obtain the "best" aperture. Automatic elliptical/circular apertures could include extra contributions from the nearby sources or miss the faint parts of the irregular galaxies.  Sometimes, it is also difficult to de-blend the flux from the nearby objects with an elliptical or circular aperture in a crowded region.  The uncertainties in the flux measurement not only stem from the background noise but also from the shape and size uncertainties of the integrating aperture itself.

Ideally, the sky noise could be minimized and the contamination from the nearby sources could be avoided if the aperture used is similar to or has the same shape as the source.  Isophotal technique has been applied for similar purposes \citep[]{1984amd..conf..277K}.  However, isophotal apertures might miss the faint edges of the objects and thus could underestimate the fluxes from the objects; a corrected isophotal aperture photometry has thus been proposed to correct the problem \citep[]{1990MNRAS.246..433M}.  The corrected isophotal aperture assumes that the faint edge of an object is mainly affected with Gaussian atmospheric blurring; therefore, the total flux owing to the object can be derived from the isophotal aperture with a Gaussian blurring correction.  The correction might be suitable for point sources; however, it has large uncertainty when applied to spheroidal galaxies \citep[]{1996A&AS..117..393B}.   Moreover, the wing of atmospheric point spread function was found to behave more like a power law rather than a Gaussian profile \citep[]{1996PASP..108..699R}.

For merging galaxies, which usually have more complex morphology, we would likely have too many unwanted background noises or contamination sources when applying elliptical apertures. Isophotal technique might be quite suitable for merging galaxies, but it might miss the faint structures of the merging galaxies. Currently, there is no suitable aperture technique that is specifically designed for merging galaxies. To achieve this goal, we provide a completely different algorithm by utilizing pattern recognition routines for obtaining an adaptive aperture which is enlarged from the original shape of the source by extending its boundary.  We considered such an enlarged aperture is "quasi-homomorphic" to the original shape of the source.  Such a quasi-homomorphic aperture is very suitable for estimating the total magnitudes of irregular galaxies or normal sources in crowded regions.

In this Supplement, we present a new technique for obtaining adaptive homomorphic apertures automatically by using pattern recognition routines. In Section \ref{method}, we describe the adaptive aperture photometry process in detail. This technique was applied to the merging galaxies observed in several survey projects, and the results are shown in Section \ref{Application}. Finally, in the last Section, we discuss our results and summarize our methods.

\section{METHODOLOGY}\label{method}

We developed a new algorithm for measuring the magnitudes of irregular objects by using adaptive homomorphic aperture photometry.  The algorithm uses pattern recognition routines to create homomorphic apertures that are similar to the shapes of the irregular objects.  The shapes of the sources and the apertures are identified in binary images, which are converted from the original CCD images based on detection thresholds.  Our photometry processes include background subtraction, binary image creation and homomorphic dilations.  The software for these processes was implemented in the Interactive Data Language (IDL).

\subsection{Background Subtraction}

Before measuring the flux of a source in a CCD image, the background has to be subtracted from the CCD image.  A global background value is usually not representative of the entire CCD image.  To do the background subtraction, we basically follow the  background subtraction steps in the SExtractor package \citep[]{1996A&AS..117..393B}.  Our algorithm first divides the CCD image into several grids (usually 200x200 pixels for one grid) in order to get the local background values.  After rejecting the outliers, the algorithm estimates the mode ($Mode = 3 \times Median - 2 \times Mean$) and the standard deviation for each grid.  The algorithm applies a 3x3 median filter on the grids to reduce the overestimation around bright objects.  In other words, for each grid the algorithm selects the median value of its eight neighbors to replace the value of that grid.  The algorithm applies the bi-cubic-convolution interpolation to the entire CCD image to construct a smooth background map and a standard deviation map for background subtraction and following processes.  

\subsection{Binary Image Process}

The algorithm sets a threshold on the background-subtracted image to create a binary image.  A specific pixel value will be set to 1(0) if the corresponding pixel of the background-subtracted image has a value higher (lower) than the local threshold.  In the algorithm, the threshold is a free parameter, measured in units of local standard deviation $\sigma_{\mathrm{sky}}$.  The default threshold is 1.5 $\sigma_{\mathrm{sky}}$ in order to retain the faint structures of the sources.  In the binary images, a group of connected pixels that exceed the threshold value is considered as an object.

\begin{figure}                    
\centering
\includegraphics[width=0.8\textwidth]{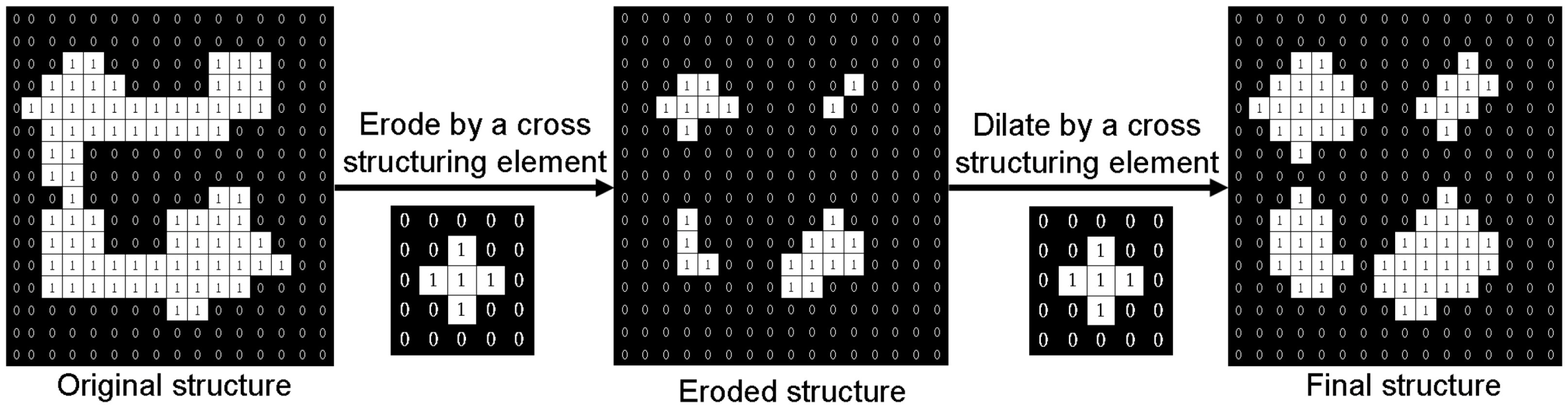}
\caption{Illustration of the OPENING operation (combination of EROSION and DILATION operators).  The result of the OPENING is shown for a cross-structure element.}
\label{fig1}
\end{figure}

To identify real sources, the algorithm applies the smoothing operator {\it Opening} on the binary image.  The operator {\it Opening} is a combination operator of the image operators {\it Erosion} and {\it Dilation}, which are two basic image operators in the area of mathematical morphology \citep[]{Pierre2003}.  The effect of {\it Dilation}/{\it Erosion} on a binary image is to enlarge/reduce the boundaries of the objects.  Graphical illustrations of these processes are shown in Figure~\ref{fig1}.   The operator {\it Opening} will eliminate the objects that are due to background noise fluctuations; these fake sources are usually have the sizes less than 5 pixels in the binary images.  Such a smoothing process might not reduce some noises that are correlated, but they will not affect on the photometry of real sources.  The {\it Opening} operator also disconnects the weak links between the object and nearby sources in the binary image.  After completing the {\it Opening}, the remaining group of connected pixels will be considered as a real object, and its coordinates will be set at the geometric center of the connected pixels.

\subsection{Homomorphical Dilations}

The adaptive aperture uses an irregular aperture that is quasi-homomorphic to the real shape of a source.  For an irregular source, the source's shape can be derived from the binary image.  However, galaxies have fluxes that decrease slowly as a function of radius and could have parts of their fluxes mixed with background noise, especially on the edges.  To capture most of the fluxes of an irregular object, our algorithm takes the object's shape in the binary image to be its original shape and increases the aperture size in a quasi-homomorphic manner, implying that the aperture's size is enlarged while trying to preserve the original source's shape.  The quasi-homomorphic extension of the aperture is implemented by using the {\it Dilation} operator.  The {\it Dilation} operator can be applied more than once for obtaining the best estimate of the flux.

The 'best' size of the aperture should be the smallest size that captures all of the source's flux.  For an isolated normal galaxy, this technique should yield the same results as other aperture techniques.  We selected several known normal galaxies, which have well-defined photometric information, for examining the relation between the measured fluxes and the aperture sizes for several different instruments.  The sizes of the apertures could be determined by comparing our results with the results for these known galaxies.  Because of the different sensitivity and observation conditions at different observations, the best aperture size for different instruments might be different.  The number of {\it Dilations} was set as a free parameter in the algorithm, for determining suitable aperture sizes for different observational data.

\section{APPLICATION}\label{Application}

To look for suitable parameters of the aperture sizes, we created 100,000 virtual galaxy images for testing.  The artificial galaxies were created by using the combination of the S$\acute{e}$rsic and the exponential profiles.  These galaxies have different bulge-to-disk ratios, S$\acute{e}$rsic indexes, and ellipticities and have the brightness difference as large as ten magnitudes.  Besides, 20,000 of the galaxies have asymmetric structures produced by adding an additional S$\acute{e}$rsic model. 

We took these 100,000 artificial galaxy images to exam the performance of our algorithms.  The results show that our method can detect more than 90\% of the total fluxes for most of the galaxies using the threshold of 1.5~$\sigma_{\mathrm{sky}}$.  These apertures with more {\it Dilations} will recover more fluxes, but had little improvement after 3 {\it Dilations}.  The testing also showed that the performance of this method is mainly affected by the the signal-to-noise ratios of the sources. 

For testing the proposed method in real data, we applied the algorithm on the Sloan Digital Sky Survey \citep[SDSS]{2000AJ....120.1579Y} images and the Red Sequence Survey 2 \citep[RCS2]{2000AJ....120.2148G} images.  We selected a group of known isolated normal galaxies for determining suitable parameters for the images.  We then applied the method by using these parameters for measuring the magnitudes of merging galaxies.  

\begin{figure}                    
\centering
\includegraphics[width=0.4\textwidth]{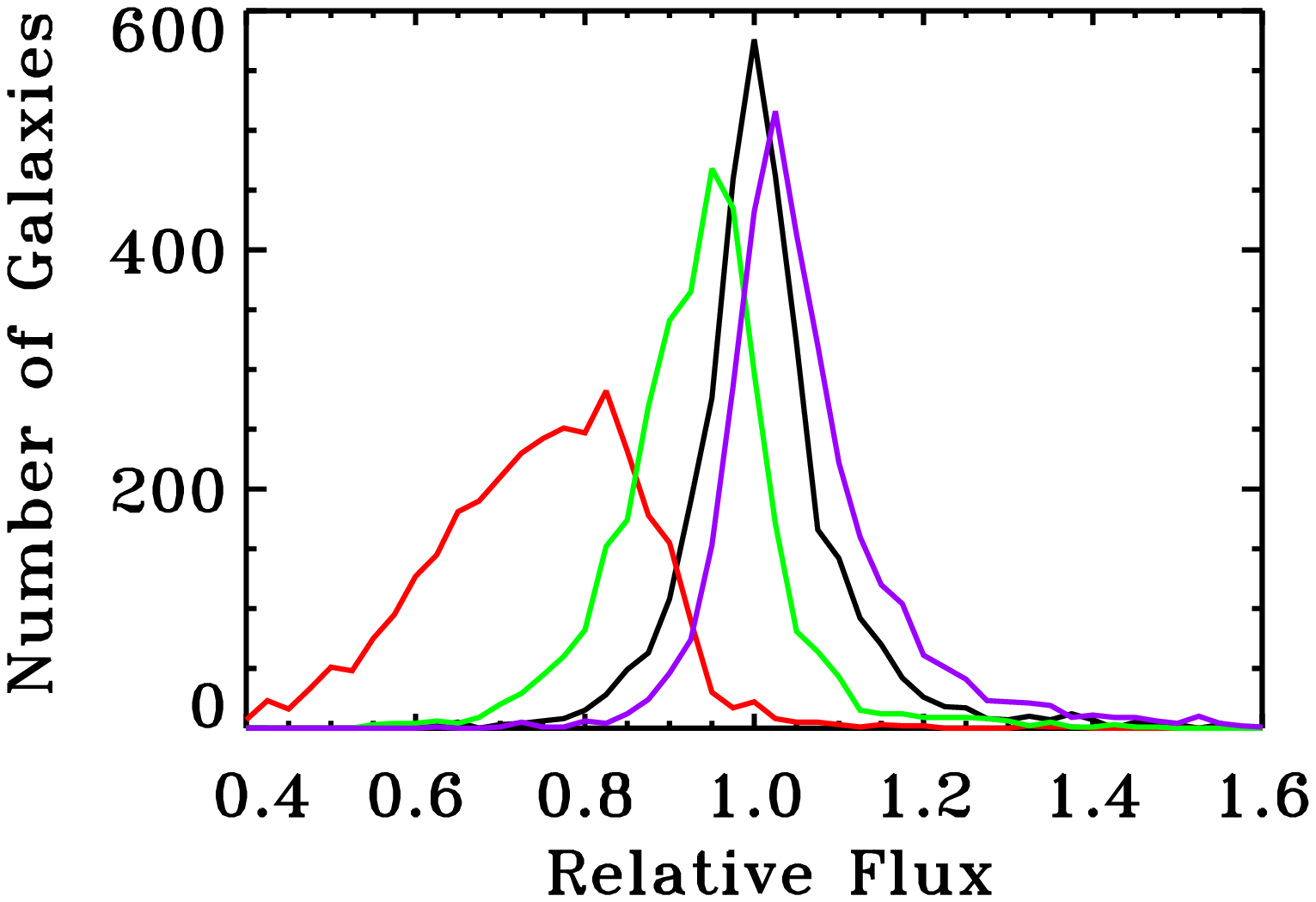}
\includegraphics[width=0.4\textwidth]{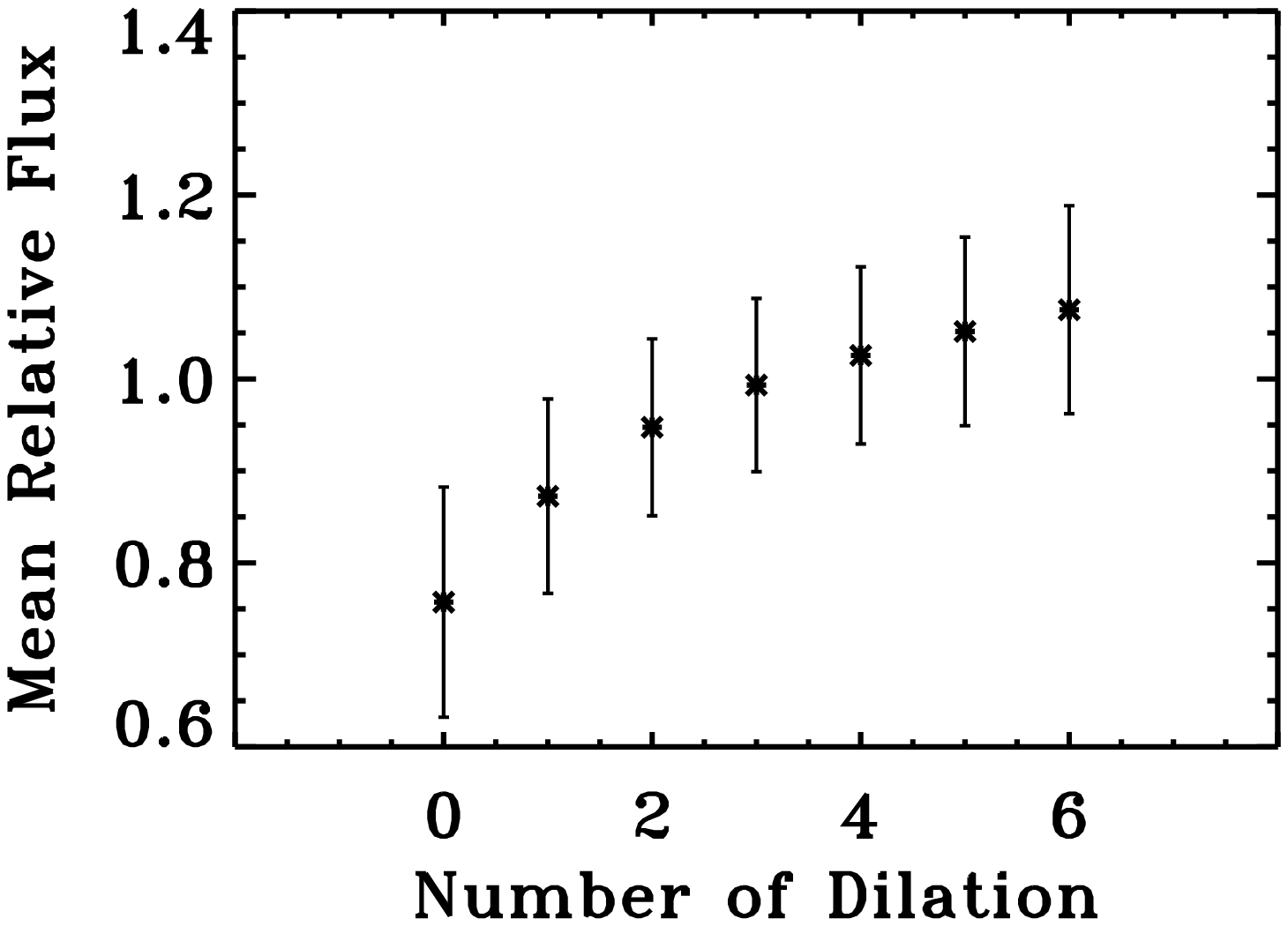}
\caption{\it Left: \rm Measured flux ratio distributions of 3,219 galaxies from the SDSS images with different aperture sizes (different dilations).  The measured flux ratio is defined as the ratio of the measured adaptive aperture flux and the SDSS Petrosian's flux.  The red, green, black, and purple lines represent the photometric results measured by using the adaptive apertures with 0, 2, 4, 6 homomorphic dilations.  \it Right: \rm Mean flux ratios for different aperture sizes. The error bars show the standard deviation of the measurements. }
\label{fig2}
\end{figure}

\begin{figure}                    
\centering
\includegraphics[width=0.6\textwidth]{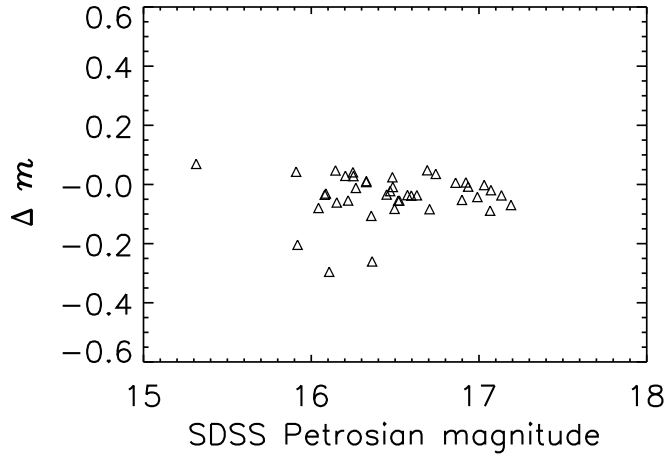}
\caption{Comparison between the magnitudes derived from the adaptive homomorphic aperture photometry and the SDSS Petrosian's photometry of 42 merging galaxies from the Galaxy Zoo.  The $\Delta m$ is defined as the adaptive homomorphic aperture magnitude minuses the SDSS Petrosian magnitude.}
\label{fig3}
\end{figure}

The SDSS Database provides well-calibrated background-subtracted images.  To test our algorithms, we selected 3,219 isolated normal galaxies from the SDSS Data release 9 and 42 mergers from the Galaxy Zoo project \citep[]{2011MNRAS.410..166L}.  Figure~\ref{fig2} shows the measured fluxes vs. the aperture sizes of the 3,219 isolated galaxies.  The results indicate that the best aperture sizes require four {\it Dilation} operations with the threshold of 1.5 $\sigma_{\mathrm{sky}}$.  We applied the parameters to obtain the homomorphic apertures for the 42 mergers and measured their fluxes.  We compared our results with the SDSS Pertrosian measurements.  Since we only measured the total fluxes of the merging galaxy systems, we summed the SDSS Pertrosian fluxes of individual galaxies in the systems for comparison.  As shown in Figure~\ref{fig3}, our algorithm was able to detect more fluxes from the merging galaxy systems. This result are most likely caused by the faint structures of the objects which did not contained in the individual galaxies measurements.  The results indicate that this method is able to detect fluxes from the faint structures of the merging galaxy systems.

\begin{figure}                    
\centering
\includegraphics[width=0.4\textwidth]{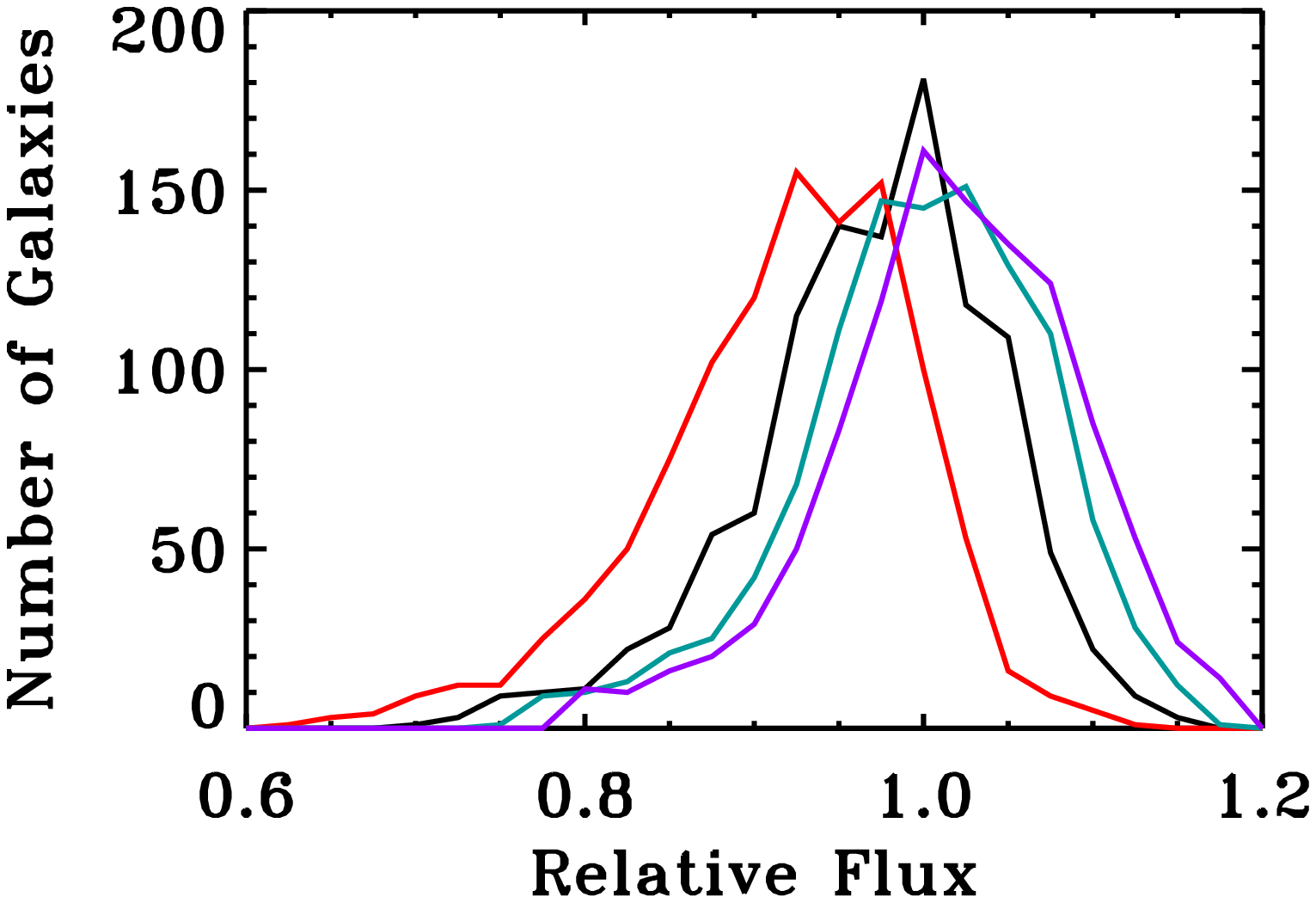}
\includegraphics[width=0.4\textwidth]{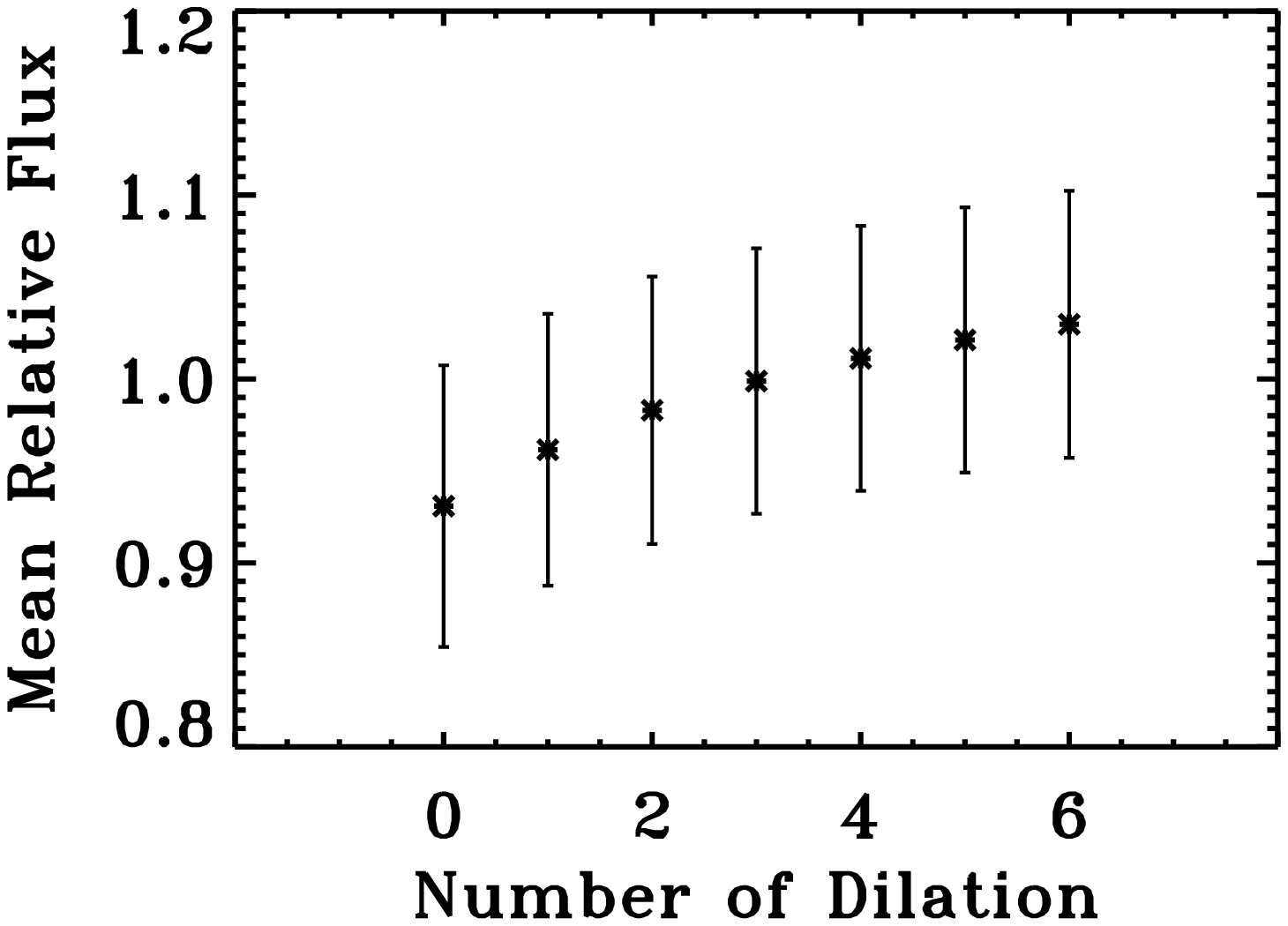}
\caption{\it Left: \rm Measured flux ratio distributions of 1,081 galaxies from the RCS2 images with different aperture sizes (different dilations).  The measured flux ratio is defined as the ratio of the measured adaptive aperture flux and the SDSS Petrosian's flux.  The red, black, green, and purple lines represent the photometric results measured by using the adaptive apertures with 0, 2, 4, 6 homomorphic dilations.  \it Right: \rm Mean flux ratios for different aperture sizes. The error bars show the standard deviation of the measurements. }
\label{fig4}
\end{figure}

\citet[]{2009ApJS..181..233H} have found more than 13,000 merging galaxy candidates by analyzing the RCS2 images.  The RCS2 images were observed by the Canada-France-Hawaii telescope (CFHT) with the wide-field imager 'MegaCam' \citep[]{2004PASP..116..449M}.  We adopted the adaptive homomorphic aperture method for retrieving the photometric information of these merging galaxy candidates.  We used 1,081 isolated normal galaxies for estimating the best dilation number in determining the size of the homomorphic apertures (Figure~\ref{fig4}).  The empirical test shows that the best size of an aperture requires 2 homomorphic dilations.  With these parameters, the photometric information on the g', r', z' bands for 13,290 mergers out of the 13,577 candidates has been successfully obtained.  For the 287 failure cases, 102 were owing to the pollution of cosmic rays; 110 sources had objects in immediate neighborhood, which were less than 0.6 arcseconds ($\sim$ 3 pixels) away from the edges of the sources; the rest were owing to false background subtraction.

\section{DISCUSSION}\label{Discussion}

We developed an adaptive homomorphic aperture technique for measuring the fluxes of irregular objects.   This technique needs to decide suitable parameters in determining the best size of an aperture.  We can use a catalog of artificial sources or a group of known isolated galaxies with regular shapes to decide the suitable parameters.  To contain minimum noises, the adaptive homomorphic apertures should be as small as possible but would capture the fluxes of the sources as much as possible.  For irregular objects, this method might also contain faint structures of the sources, which might not be included by other automatically aperture techniques.  Our test showed that our algorithms could capture more fluxes for merging galaxies than did the Petrosian aperture technique (Figure~\ref{fig3}), whereas these two techniques demonstrated similar results for normal galaxies.  This results indicate that the adaptive homomorphic apertures can be useful for detecting faint structures, which might not be detected by using the Petrosian aperture technique.

In the algorithms, the threshold and the number of {\it Dilations} are two main parameters that need to be decided.  Threshold value will determine the shapes of the objects. Using high threshold values can minimize the contamination of background noises and separate very nearby sources but may loss the structure information at the edges of the objects.    Through the test for the known isolated galaxies with regular shapes, we found that the threshold value and the number of {\it Dilations} are not independent; using high threshold values need more {\it Dilations} in order to have more accurate estimation of the fluxes of the sources.

The adaptive homomorphic aperture technique requires different number of {\it Dilations} for yielding the best aperture sizes for images with different sensitivity and resolutions.  Our tests showed that the number of {\it Dilations} for the SDSS images is four; however, under the same threshold condition, the number of {\it Dilations} for the RCS2 images is two.  The method might require a group of known isolated galaxies with regular shapes for empirically deciding the suitable parameters for different observations.  The influence of seeing will also be taken into account through the parameter deciding; The dilated regions usually need to be greater than or similar to the seeing.

The adaptive homomorphic aperture technique is also useful for objects located in the crowded regions.  Many merging galaxies in the RCS2 images were detected with additional nearby objects. The median seeing of the RCS2 images is about 0.7 arcseconds and the pixel sizes are about 0.2 arcseconds.  With these observational conditions, we found that the homomorphic apertures can avoid contaminations from the nearby objects as long as the nearby objects are more than 0.6 arcseconds away from the edges of the sources.

The images in different observing bands may require different parameters in creating the best adaptive homomorphic apertures.  This might cause a problem when estimating the colors of objects since the detected areas on different images are different.  To avoid this problem, we used a common aperture to measure the fluxes of an object in different observing bands.  The basic shape in creating the common aperture of an object was the union of three basic shapes from aligned images of different observing bands. 

This technique is similar to the isophotal technique \citep[]{1984amd..conf..277K, 1996A&AS..117..393B}.  The isophotal aperture photometry determines the aperture sizes by considering the pixels that have signals above the detection threshold, which is similar to the sizes of the original objects in the first step of the binary image process in our algorithm.  Generally speaking, the isophotal technique might miss the faint structures of the objects and might underestimate their total fluxes.  There is a corrected isophotal aperture photometry attempting to correct the problem theoretically \citep[]{1990MNRAS.246..433M}.  The corrected isophotal apertures assume that the faint edges of the objects mainly stem from the Gaussian atmospheric blurring; thus, the total fluxes of the objects can be derived from the isophotal aperture photometry results by applying the correction relation.  On the other hand, our algorithm avoids this problem by using the homomorphic dilation process.  We applied these three aperture photometry techniques on the isolated normal galaxy samples of the SDSS and the RCS2 for examining their performance (Figure~\ref{fig7}).  We found that the measurement results obtained by using our algorithm are comparable with the results of the corrected isophotal technique for the RCS2 galaxies; however, for the SDSS sources, the results obtained by our algorithm are consistent with those of the Petrosian measurement results, whereas both the isophotal and corrected isophotal measurement results seem to underestimate the true fluxes.  These results suggest that the adaptive homomorphic aperture technique works well for different observations and the photometry information can be acquired in a straightforward manner. 

The adaptive homomorphic aperture technique is applicable to irregularly shaped extended sources and is also useful for sources in crowded regions.  However, it might be necessary to decide the best number of dilations empirically for different instruments.  Once the number of dilations has been decided for specific observation data, the algorithms can be used for obtaining the fluxes of the sources.

\section*{Acknowledgments}

This work was partially supported by the Ministry of Science and Technology (MOST) of Taiwan through grant MOST 103-2119-M-008-017-MY3.

\clearpage

\begin{figure}                    
\centering
\includegraphics[width=0.5\textwidth]{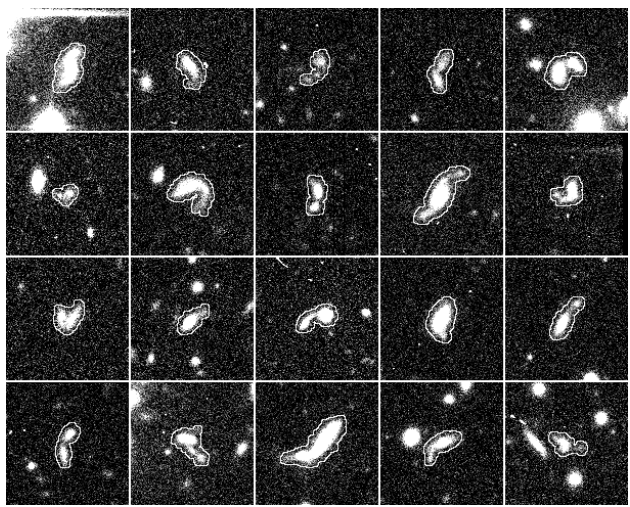}
\caption{Examples of adaptive homomorphic apertures for 20 merging galaxies from the RCS2 r' band images.  The apertures were selected to include the whole merging galaxy systems.} 
\label{fig5}
\end{figure}

\begin{figure}                    
\centering
\includegraphics[width=0.4\textwidth]{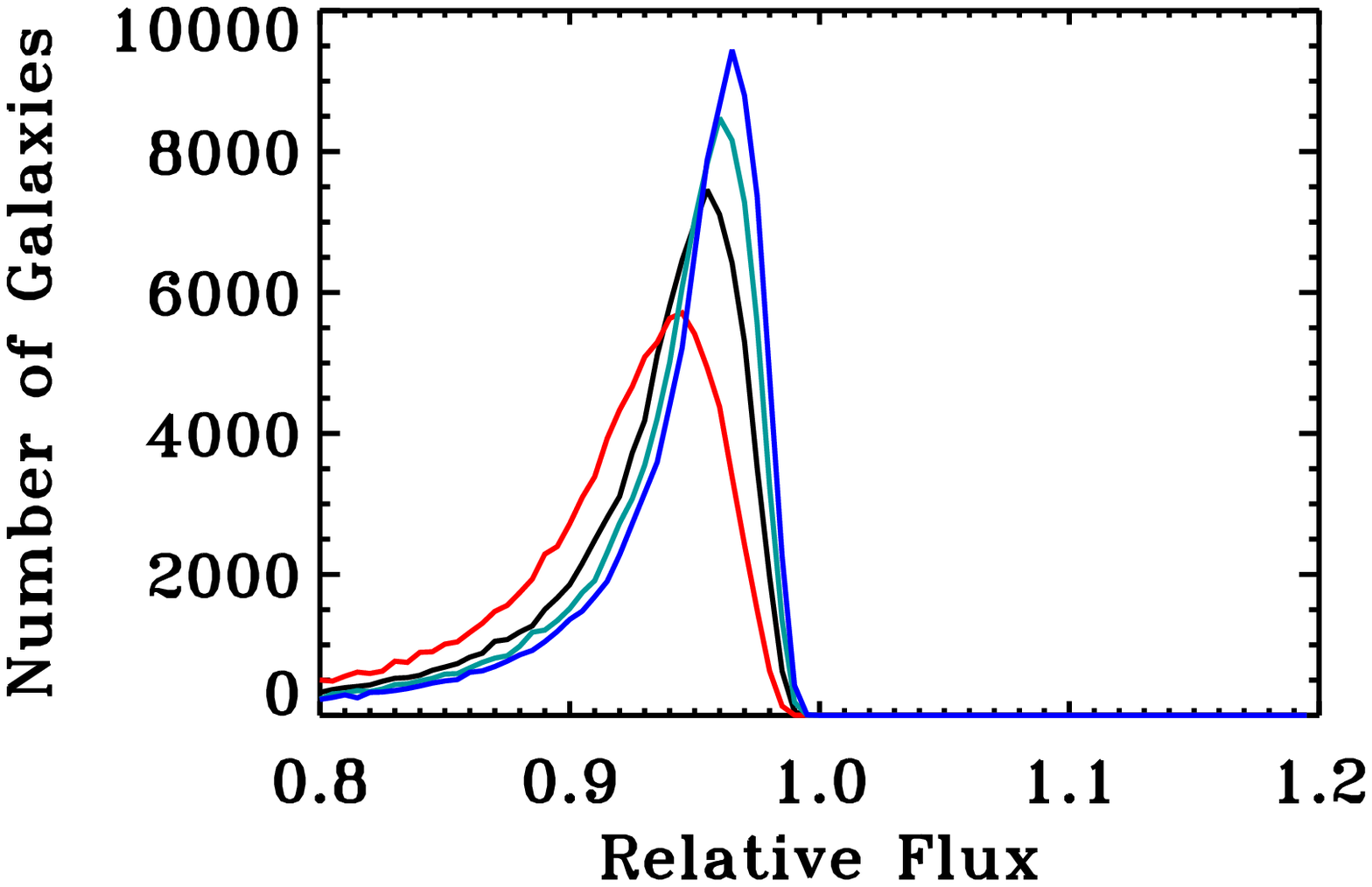}
\includegraphics[width=0.4\textwidth]{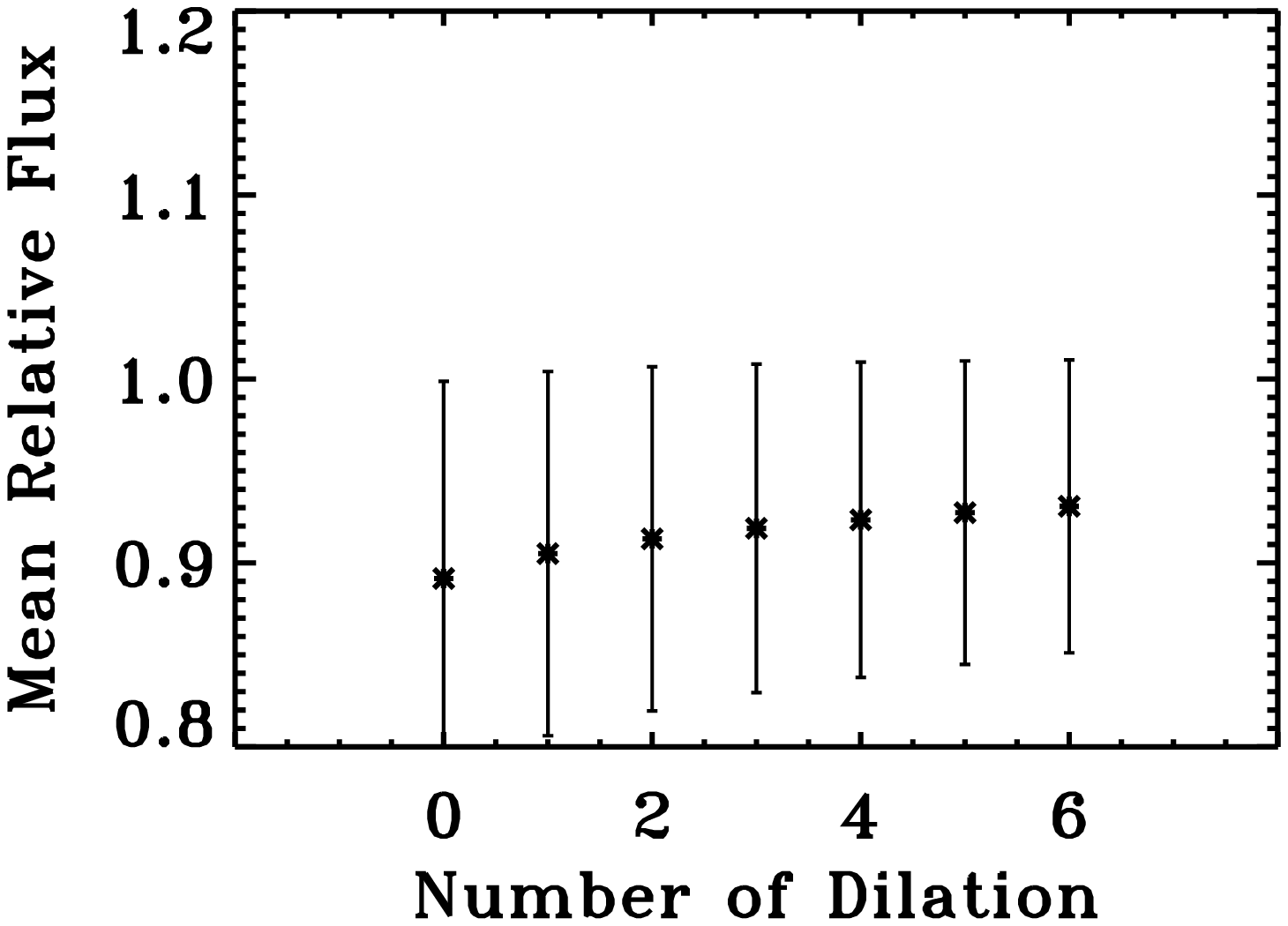}
\caption{\it Left: \rm Measured flux ratio distributions from 100,000 artificial galaxy images with different aperture sizes (different dilations).  The measured flux ratio is defined as the ratio of the measured adaptive aperture flux and the theoretical flux.  The red, black, green, and purple lines represent the photometric results measured by using the adaptive apertures with 0, 2, 4, 6 homomorphic dilations.  \it Right: \rm Mean flux ratios for different aperture sizes.  The error bars show the standard deviation of the measurements.} 
\label{fig6}
\end{figure}

\begin{figure}                    
\centering
\includegraphics[width=0.4\textwidth]{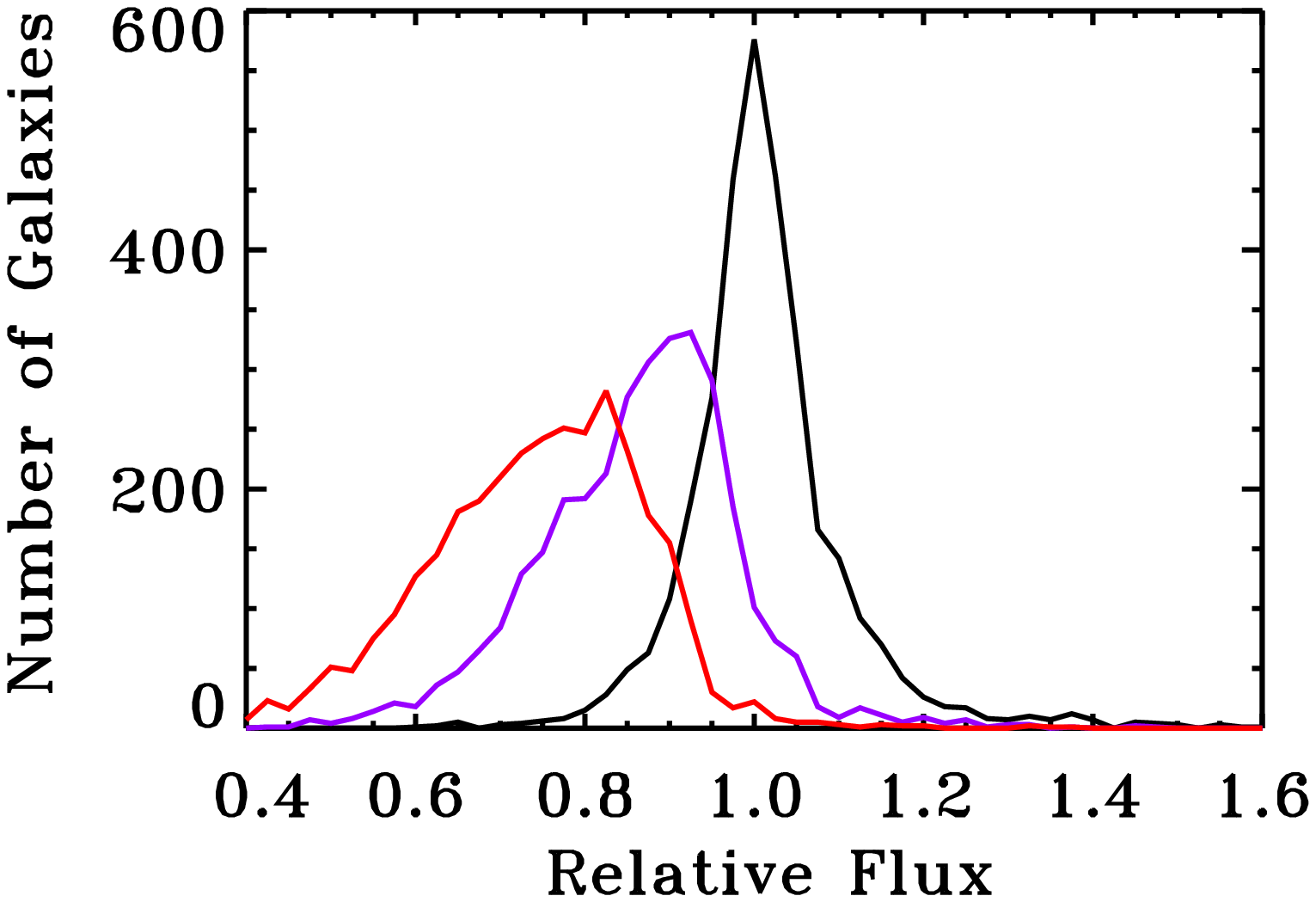}
\includegraphics[width=0.4\textwidth]{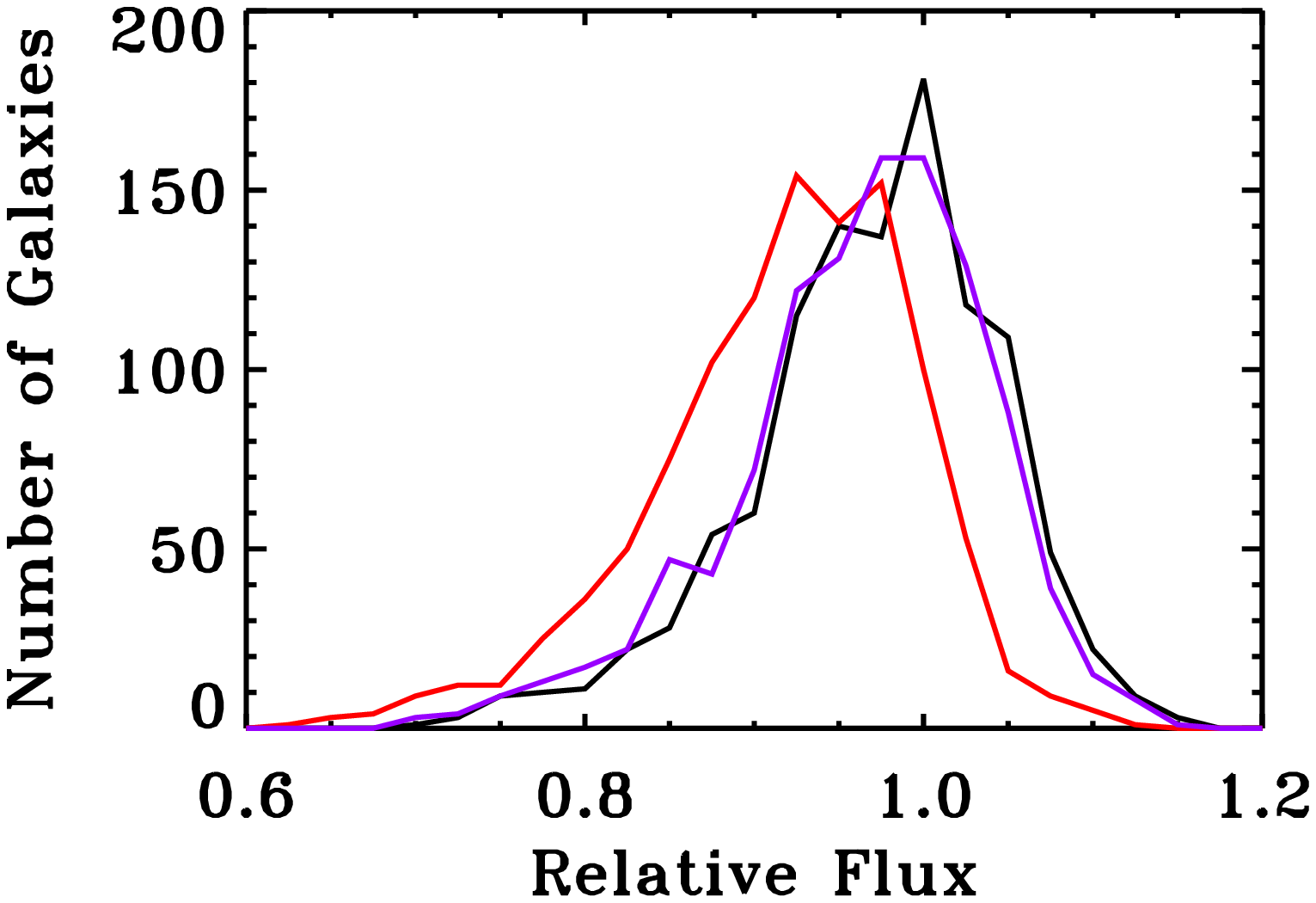}
\caption{\it Left: \rm Measured flux ratio distributions of 3,219 galaxies in the SDSS images by using different aperture methods.  \it Right: \rm Results for the 1,081 galaxies in the RCS2 images.  The measured flux ratio is defined as the ratio of the measured flux and the SDSS Petrosian flux. The red lines are the results for the isophotal apertures, the purple lines are the results for the corrected isophotal apertures, and the black lines are the results for the adaptive homomorphic apertures. }
\label{fig7}
\end{figure}


\begin{thebibliography}{99}
\bibitem[\protect\citeauthoryear{Bertin \& Arnouts}{1996}]{1996A&AS..117..393B} Bertin E., Arnouts S., 1996, A\&AS, 117, 393
\bibitem[\protect\citeauthoryear{Gladders \& Yee}{2000}]{2000AJ....120.2148G} Gladders M.~D., Yee H.~K.~C., 2000, AJ, 120, 2148 
\bibitem[\protect\citeauthoryear{Hwang \& Chang}{2009}]{2009ApJS..181..233H} Hwang C.-Y., Chang M.-Y., 2009, ApJS, 181, 233 
\bibitem[\protect\citeauthoryear{Kibblewhite et al.}{1984}]{1984amd..conf..277K} Kibblewhite E.~J., Bridgeland M.~T., Bunclark P.~S., Irwin M.~J., 1984, amd..conf, 277 
\bibitem[\protect\citeauthoryear{Kron}{1980}]{1980ApJS...43..305K} Kron R.~G., 1980, ApJS, 43, 305 
\bibitem[\protect\citeauthoryear{Lintott et al.}{2011}]{2011MNRAS.410..166L} Lintott C., et al., 2011, MNRAS, 410, 166 
\bibitem[\protect\citeauthoryear{Maddox, Efstathiou, \& Sutherland}{1990}]{1990MNRAS.246..433M} Maddox S.~J., Efstathiou G., Sutherland W.~J., 1990, MNRAS, 246, 433
\bibitem[\protect\citeauthoryear{Magnier \& Cuillandre}{2004}]{2004PASP..116..449M} Magnier E.~A., Cuillandre J.-C., 2004, PASP, 116, 449 
\bibitem[\protect\citeauthoryear{Petrosian}{1976}]{1976ApJ...209L...1P} Petrosian V., 1976, ApJ, 209, L1
\bibitem[\protect\citeauthoryear{Racine}{1996}]{1996PASP..108..699R} Racine, Rene, 1996, PASP, 108, 699R
\bibitem[\protect\citeauthoryear{Pierre}{2003}]{Pierre2003} Pierre Soille, 2003, Morphological Image Analysis: Principles and Applications. 2nd edition. Springer-Verlag, Berlin
\bibitem[\protect\citeauthoryear{York et al.}{2000}]{2000AJ....120.1579Y} York D.~G., et al., 2000, AJ, 120, 1579 
\end{thebibliography}
\end{document}